\documentclass[sigconf,natbib=true]{acmart}
\usepackage{graphicx} % DO NOT CHANGE THIS
\usepackage{caption}
\usepackage[linesnumbered,ruled,vlined]{algorithm2e}
\usepackage{subfigure}
\usepackage{multirow}
\usepackage[normalem]{ulem}
\usepackage{balance}
\usepackage[T1]{fontenc}
\usepackage[utf8]{inputenc}
\usepackage{makecell}
\usepackage{enumitem}
\usepackage{amsmath}
\usepackage{url}

\SetKwInput{KwInput}{Input}                % Set the Input
\SetKwInput{KwOutput}{Output}              % set the Output

\AtBeginDocument{%
  \providecommand\BibTeX{{%
    \normalfont B\kern-0.5em{\scshape i\kern-0.25em b}\kern-0.8em\TeX}}}

\copyrightyear{2023}
\acmYear{2023}
\setcopyright{acmlicensed}\acmConference[ORSUM '23]{ORSUM@ACM RecSys 2023, 6th Workshop on Online Recommender Systems and User Modeling, jointly with the 17th ACM Conference on Recommender Systems}{September 19th, 2023}{Singapore}
\acmBooktitle{ORSUM@ACM RecSys 2023, 6th Workshop on Online Recommender Systems and User Modeling, jointly with the 17th ACM Conference on Recommender Systems, September 19th, 2023, Singapore}
\acmPrice{15.00}
\acmDOI{XXXXXXX.XXXXXXX}
\acmISBN{978-1-4503-XXXX-X/23/06}

\begin{document}

\title[Deep Mutual Learning for Effective Multi-Task Recommender Learning]{Deep Mutual Learning across Task Towers for Effective Multi-Task Recommender Learning}


\author{Yi Ren}
\affiliation{%
  \institution{Tencent }
  \city{Beijing}
  \country{China}}  
\email{yiren_bj@outlook.com}

\author{Ying Du}
\affiliation{%
  \institution{Tencent}
  \city{Beijing}
  \country{China}}  
\email{yingdu@tencent.com}
% \footnotemark[\value{footnote}]

\author{Bin Wang}
\affiliation{%
  \institution{Tencent}
  \city{Beijing}
  \country{China}}  
\email{hillmwang@tencent.com}
% \footnotemark[\value{footnote}]

\author{Shenzheng Zhang}
\affiliation{%
  \institution{Tencent }
  \city{Beijing}
  \country{China}}    
\email{qjzcyzhang@tencent.com}

\renewcommand\shortauthors{Yi Ren et al.}

\begin{abstract}
Recommender systems usually leverage multi-task learning methods to simultaneously optimize several objectives because of the multi-faceted user behavior data. The typical way of conducting multi-task learning(MTL) is to establish appropriate parameter sharing across multiple tasks at lower layers while reserving a separate task tower for each task at upper layers. With such design, the lower layers intend to explore the structure of task relationships and mine valuable information to be used by the task towers for accurate prediction. 

Since the task towers exert direct impact on the prediction results, we argue that the architecture of standalone task towers is sub-optimal for promoting positive knowledge sharing. First, for each task, attending to the input information of other task towers is beneficial. For instance, the information useful for predicting the "like" task is also valuable for the "buy" task. Furthermore, because different tasks are inter-related, the training labels of multiple tasks should obey a joint distribution. It is undesirable for the prediction results for these tasks to fall into the low density areas. Accordingly, we propose the framework of \textbf{D}eep \textbf{M}utual \textbf{L}earning across task towers(\textbf{DML}), which is compatible with various backbone multi-task networks. At the entry layer of the task towers, the shared component of Cross Task Feature Mining(\textit{CTFM}) is introduced to transfer input information across the task towers while still ensuring one task’s loss will not impact the inputs of other task towers. Moreover, for each task, dedicated network component called Global Knowledge Distillation(\textit{GKD}) are utilized to distill valuable knowledge from the global results of the upper layer task towers to enhance the prediction consistency. Extensive offline experiments and online A/B tests are conducted to evaluate and verify the proposed approach's effectiveness.

\end{abstract}




\maketitle

%\vspace{-0.1cm}

\section{Introduction}
Recently, we have seen the widespread application of recommender systems, which involve different types of user feedback signals, such as clicking, rating, commenting, etc. Moreover, no single feedback signal can accurately reflect user satisfaction. For example, over-concentrating on clicking may aggravate the click-bait issue. Therefore, it is highly desirable to be able to effectively learn and estimate multiple types of user behaviors at the same time. And Multi-Task Learning is a promising technique to address this challenge. Given several related learning tasks, the goal of multi-task learning is to enhance the overall performance of different tasks by leveraging knowledge transfer among tasks. With the Multi-Task Learning(MTL) paradigm\cite{zhang2021survey,caruana1997multitask}, multiple tasks are learned simultaneously in a single model. Compared with single-task solutions, MTL costs much fewer machine resources and improves learning efficiency since we just need to train and deploy a single model. Moreover, MTL usually can warrant enhanced recommendation performance through appropriate parameter sharing. 

Most existing methods of Multi-Task Learning(MTL) appropriately share parameters across multiple tasks at lower layers while keeping separate task towers at upper layers. These methods can be roughly classified into four categories. The first category comprises the methods of \textbf{\textit{hard parameter sharing}}. Among them, embedding sharing is the most intuitive structure to share information. For instance, the ESSM model \cite{ma2018entire} shares embedding parameters between the tasks of CTR (Click-Through Rate) and CVR (Conversion Rate) for improving the prediction performance of the sparse CVR task. In addition to the embedding parameters, the Shared-Bottom structure\cite{caruana1997multitask} is introduced to share the parameters of lower-layer MLPs among tasks. But these methods are severely plagued by the task conflicts and negative transfer issue. Second, for the methods of \textbf{\textit{soft parameter sharing}}, each task owns separate parameters, which are regularized during training to minimize the differences between the shared parameters. L2-constrained\cite{duong2015low} is a typical algorithm belonging to this category.  Third, for the methods of \textbf{\textit{customized routing}}, they learn customized routing weights for each task to combine and fuse information from lower-layer networks to counteract the negative transfer issue. Cross-stitch network\cite{misra2016cross} and sluice network\cite{ruder2019latent} learn separate linear weights for each task to selectively merge representations from different lower-level branches. SNR\cite{ma2019snr} modularizes the shared low-level layers into parallel sub-networks and uses a transformation matrix multiplied by a scalar coding variable to learn their connections to upper-level layers to alleviate the task conflict and negative transfer issue. MSSM\cite{ding2021mssm} learns differing combinations of feature fields for each expert and designs finer-grained sharing patterns among tasks through a set of coding variables that selectively choose which cells to route for a given task. But the learned routing parameters of these methods are static for all the samples, which can hardly warrant optimal performance. Finally, the methods of \textbf{\textit{dynamic gating}} learn optimized weights for each task based on the input sample to effectively combine the outputs of lower-level networks and achieve success in industrial applications. The MMoE \cite{ma2018modeling} model adapts the Mixture-of-Experts (MoE)\cite{jacobs1991adaptive} structure to multi-task learning by sharing the expert sub-networks across all tasks, while also maintaining separate gating network optimized for each task. And Zhao et al. \cite{zhao2019recommending} extend the MMoE model \cite{ma2018modeling} and apply it to learn multiple ranking objectives in Youtube video recommender systems. Moreover, PLE \cite{tang2020progressive} achieves superior performance for news recommendation by assigning both shared expert sub-networks among tasks and dedicated expert sub-networks for each task. 

AITM \cite{xi2021modeling} is the most similar method, which also augments the arcitecture of task towers. Nevertheless, as a concrete implementation, it is not validated to enhance the performance of various multi-task models. Moreover, it can only work for the tasks with sequential dependence relations. 

Admittedly, the aforementioned methods achieve impressive performance. However, as the task towers exert a direct effect on the prediction results, the standalone task towers tend not to be the most effective design for promoting positive knowledge transfer by exploiting the task relationships. First, for each task, the information selected by the relevant tasks is extremely valuable. Accordingly, we introduce the shared component of Cross Task Feature Mining(\textit{CTFM}), which utilizes delicate attention mechanisms to extract relevant information from other tasks at the entry layer of the task tower. With the common attention mechanisms, the explicit task-specific information distilled by lower-level networks are mingled together and one task's loss will undesirably affect the inputs of other task towers, which is the task awareness missing problem and can hinder the learning of lower-level networks. In contrast to the usual attention mechanisms, our design can ensure appropriate information separation. We argue that reserving explicit task-specific knowledge has a positive effect on performance, which is validated in the experimental section. Second, because the tasks for recommender systems are related, the training labels of multiple tasks should obey a joint distribution. The prediction results for these tasks should not densely fall into the low-density areas. Therefore, a dedicated network named Global Knowledge Distillation(\textit{GKD}) is introduced for each task to distill valuable global knowledge from the results of the upper layer task towers. For each task, the distilled global information helps to ensure consistent predictions with other tasks. We summarize our main contributions below.   
%\vspace{-0.1cm}
\begin{itemize}[leftmargin=.2in]
\item We propose the framework of \textbf{D}eep \textbf{M}utual \textbf{L}earning across task towers(\textbf{DML}), which is compatible with various backbone multi-task models. 
\item The proposed novel sharing structure helps to enhance effective knowledge transfer across different tasks. 
\item We conduct offline experiments and online A/B testing to evaluate and understand the effectiveness of our method.
\end{itemize}

%\vspace{-0.4cm}
\section{Methodology}
In this section, we first introduce the problem of multi-objective ranking for recommender systems. Second, we describe the general design of DML. Finally, we elaborate on the introduced components.
%\vspace{-0.6cm}
\subsection{Multi-Objective Ranking for Recsys}
Given a set of candidates with $N$ items $\mathcal{C} = \{i_n\}_{1 \leq n \leq N}$, the ranking model for recommender systems is to rank and recommend the top $M$ items $\mathcal{S} = \{i_m\}_{1 \leq m \leq M} \subseteq \mathcal{C}$ for user $u$ so as to optimize the overall utility and enhance user experience. First, for each pair of user $u$ and item $i_n$, the input feature $x_n$ is derived. Second, a multi-task learning model is utilized to estimate $K$ objectives corresponding to multiple user feedback signals. Furthermore, to compute the overall reward, we need to merge the multiple predictions with a function $\Phi$ shown in equation \eqref{eq:score_merge} to derive the item's final reward score for greedy ranking. 
%\vspace{-0.1cm}
\begin{equation} \label{eq:multi_obj_pred}
o^1_n,o^2_n,...,o^K_n = MTLModel(x_n;\theta)
\end{equation}
%\vspace{-0.5cm}
\begin{equation} \label{eq:score_merge}
r_n = \Phi(o^1_n,o^2_n,...,o^K_n)
\end{equation}
%\vspace{-0.1cm}
where $\theta$ denotes the model parameters and $\Phi$ is usually a function manually tuned to reflect the reward of recommending item $i_n$ to user $u$ based on the business goals. With the estimated reward scores $\{r_n\}_{1 \leq n \leq N}$ for the items in $\mathcal{C}$, the ranking model can recommend the item sequence $\mathcal{S}$ consisting of the top $M$ items to the user $u$.
%\vspace{-0.2cm}
\begin{figure} 
  \centering 
   \includegraphics[width=2.5in]{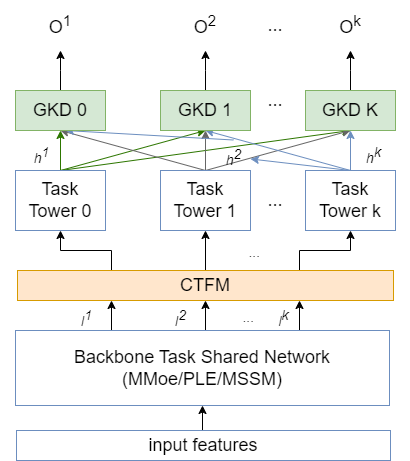} 
  \caption{Overall Model Structure of DML}
  \label{fig:dml_overall}
\end{figure}
\subsection{Overall Design of DML}
Please refer to figure \ref{fig:dml_overall} for the overall network architecture of DML. With the existing MTL algorithms\cite{zhang2021survey}, the equation \eqref{eq:multi_obj_pred} can be further decomposed as below. For simplicity, we omit the subscript $n$ in this section.
%\vspace{-0.1cm}
\begin{equation} \label{eq:encoder}
l^1,l^2,...,l^K = G(x;\theta_l)
\end{equation}
%\vspace{-0.4cm}
\begin{equation} \label{eq:single_target_score}
o^k = F^k(l^k;\theta^k_h)
\end{equation}
%\vspace{-0.1cm}
where $G$ represents the lower level networks that encodes $x$ to $K$ different latent spaces with partially or fully shared parameters $\theta_l$. And $F^k$ is the upper-level network for task $k$, which accepts $l^k$ as input to model objective $k$ with task-specific higher level parameter $\theta^k_h$. Multiple candidate models \cite{ma2018entire,ma2018modeling,tang2020progressive,ma2019snr,zhao2019recommending,misra2016cross,ding2021mssm,duong2015low,ruder2019latent,zhang2021survey} are proposed to enhance $G$ with different parameter sharing designs. 

In this research, rather than $G$, we focus on the enhancement of upper-level networks for improved prediction performance. First, the shared component of \textit{CTFM} is introduced, which leverages the attention mechanism to extract relevant information from the inputs of other task towers (the results of Equation \eqref{eq:encoder}) as a complement to the target task. Please note that this attention is well-designed to solve the task-awareness missing issue, for which the excessive encouragement of knowledge sharing is not conducive to the extraction of task-specific knowledge. With our design, the gradients computed from the target task's loss will not impact the inputs of other task towers.
%\vspace{-0.1cm}
\begin{equation} \label{eq:dml_ctfm}
\hat{l}^1,\hat{l}^2,...,\hat{l}^K = CTFM(l^1,l^2,...,l^K;\theta_s)
\end{equation}
%\vspace{-0.1cm}
where shared parameters $\theta_s$ is employed across different tasks. 

Moreover, a separate multi-layer network is introduced for each task to process each element of $\{\hat{l}^k\}_{1 \leq k \leq K}$ and generate the hidden representation, based on which accurate prediction can be made.
%\vspace{-0.1cm}
\begin{equation} \label{eq:dml_forward}
h^k = H^k(\hat{l}^k;\theta^k_{t_0})
\end{equation}
where $H^k$ denotes the task-specific MLP for task $k$ with separate parameters $\theta^k_{t_0}$.

Finally, for each task, a dedicated component named \textit{GKD} is utilized to distill information from the hidden representations for both itself and other tasks to promote prediction consistency across tasks and more precisely model the corresponding objective. 
%\vspace{-0.1cm}
\begin{equation} \label{eq:dml_gkd}
o^k = GKD^k(h^k, \{h^j\}_{1 \leq j \leq K};\theta^k_{t_1})
\end{equation}
where $GKD^k$ is the dedicated component for task $k$. In contrast to $CTFM$ at Equation \eqref{eq:dml_ctfm}, we utilize task-specific parameters $\theta^k_{t_1}$ here as specialization is usually helpful for upper layer networks. Furthermore, proper operation is implemented to ensure the prediction error of one task does not impact the hidden representations of other tasks.

\begin{figure} 
  \centering 
  \subfigure[Cross Task Feature Mining (Shared)]{ 
    \label{fig:dml:ctfm} %% label for first subfigure 
    \centering
    % \begin{minipage}[t]{0.25\linewidth}
    \includegraphics[height=1.5in]{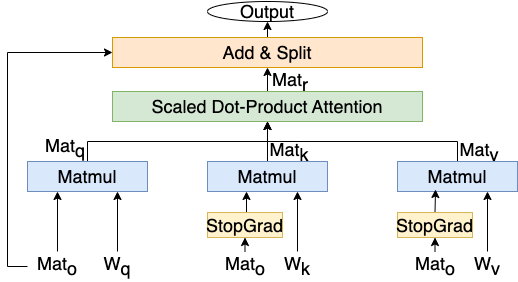} 
    % \end{minipage}
    }
  \subfigure[Global Knowledge Distillation (for Task k)]{ 
    \label{fig:dml:gkd} %% label for first subfigure 
    \centering
    % \begin{minipage}[t]{0.25\linewidth}
    \includegraphics[height=1.5in]{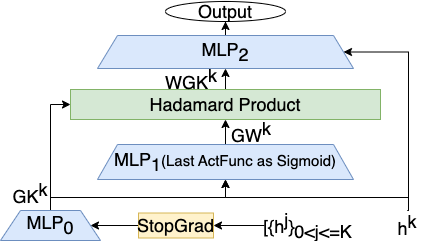} 
    % \end{minipage}
    }
    \caption{Introduced Components of DML}
    \label{fig:dml}
\end{figure}
%\vspace{-0.3cm}
\subsection{Cross Task Feature Mining}
Please refer to Figure \ref{fig:dml:ctfm} for the detailed process of \textit{CTFM}. From the perspective of task towers, the outputs from lower-level networks at Equation \eqref{eq:encoder} can be recognized as the mined features for them. First, for each task, the features mined for related tasks can be leveraged to enhance its prediction performance. Thus, trainable task embeddings, namely $\{t^k\}_{1 \leq k \leq K}$, are derived for the tasks to facilitate the learning of general task relations. Second, the importance of features from related tasks can vary per sample. Accordingly, we stack together the items of the set $\{l^k+t^k\}_{1 \leq k \leq K}$ to derive the matrix $Mat_o \in \mathcal{R}^{K \times d_0}$ where $d_0$ is the size of $l^k$ and $t^k$. Third, we use the projection parameters $W_q, W_k, W_v \in \mathcal{R}^{d_0 \times d_0}$ to transform $Mat_o$ to the query, key, and value matrix of $Mat_q, Mat_k, Mat_v \in \mathcal{R}^{K \times d_0}$. And gradient backpropagation from $Mat_k$ and $Mat_v$ to $Mat_o$ is forbidden. Moreover, the scaled dot-product attention \cite{vaswani2017attention} is performed on $Mat_q$, $Mat_k$, and $Mat_v$ to compute the result matrix $Mat_r \in \mathcal{R}^{K \times d_0}$. Finally, we add $Mat_o$ to $Mat_r$ for residual connection and further split based on the first axis to return $\{\hat{l}^k\}_{1 \leq k \leq K}$. Please note that the aforementioned networks are shared among the tasks to encourage generalizable modeling with parameter sharing.  The usual attention mechanisms will cause the task awareness missing problem and can hinder the learning of lower-level networks. Instead, our design can ensure appropriate information separation and reserve the explicit task-specific knowledge by stopping the gradient backflow from $Mat_k$ and $Mat_v$ to $Mat_o$.   
%\vspace{-0.3cm}
\subsection{Global Knowledge Distillation}
Please refer to Figure \ref{fig:dml:gkd} for the detailed process of \textit{GKD}. In contrast to \textit{CTFM}, each task is assigned dedicated parameters for global knowledge distillation. Acting as the last step, we would like to facilitate more flexible modeling by promoting task specialization here. This module accepts the hidden representations for both the corresponding task and other tasks as input. First, a multilayer perceptron(MLP) is utilized to extract valuable global knowledge (${GK}^k \in \mathcal{R}^{d_1}$) for the target task $k$ from the concatenation of all these hidden representations ($\{h^j\}_{1 \leq j \leq K}$). Since the design goal here is to train task-specific MLPs to distill relevant global knowledge while not impacting the hidden representations, the gradient backpropagation from $GK^k$ to the MLP's input is prohibited. Second, we input ${GK}^k$ and $h^k$ to another MLP with Sigmoid as its last activation function. Then, the weights of ${GK}^k$'s different latent dimensions can be dynamically adapted for each sample. The weights are denoted by ${GW}^k \in \mathcal{R}^{d_1}$. Moreover, the weighted global knowledge (${WGK}^k \in \mathcal{R}^{d_1}$) is computed with the hadamard product of ${GK}^k$ and ${GW}^k$. Finally, ${WGK}^k$ and $h^k$ are concatenated together as the input for the last MLP to make predictions for task $k$.

%\vspace{-0.2cm}
\section{Experiments}

In this section, we conduct extensive offline experiments\footnote{The code can be found at: https://github.com/renyi533/mtl-consistency/tree/main.} and online A/B testing to prove DML's effectiveness.
\begin{table}
%\footnotesize
\normalsize
\centering
\caption{The overall performance. The bold-face font denotes the winner in that column. Moreover, the "*" symbol denotes introducing DML achieves significant (p < 0.05 for one-tailed t-test) gain over the corresponding baseline. }
\label{tab:overview}
\begin{tabular}{llll|ll} 
\hline
\multirow{2}{*}{Model} & \multicolumn{3}{c|}{ML-1M}                                                                                                      & \multicolumn{2}{c}{Electronics}                    \\ 
\cline{2-6}
                       & $AUC$                   & $MSE$                   & \begin{tabular}[c]{@{}c@{}}\textit{Consistent}\\\textit{Ratio}\end{tabular} & $AUC_{rate}$                 & $AUC_{pos}$                  \\ 
\hline\hline
Single Task            & 0.8066                  & 0.7741                  & 0.7154                                                                      & 0.7608                  & 0.7334                   \\ 
\hline
SB                     & 0.8100                  & 0.7724                  & 0.7530                                                                      & 0.7876                  & 0.7608                   \\
SB+DML                 & ${0.8115}^*$          & ${0.7648}^*$          & $\boldsymbol{0.7649}^*$                                                     & ${0.7890}^*$          & ${0.7631}^*$           \\ 
\hline
MSSM                   & 0.8128                  & 0.7651                  & 0.7519                                                                      & 0.7883                  & 0.7627                   \\
MSSM+DML               & ${0.8141}^*$          & ${0.7611}^*$          & ${0.7637}^*$                                                              & ${0.7892}^*$          & ${0.7641}^*$           \\ 
\hline
MMOE                   & 0.8105                  & 0.7688                  & 0.7507                                                                      & 0.7888                  & 0.7628                   \\
MMOE+DML               & ${0.8135}^*$          & ${0.7591}^*$          & ${0.7596}^*$                                                              & $\boldsymbol{0.7897}^*$ & $\boldsymbol{0.7644}^*$  \\ 
\hline
PLE                    & 0.8122                  & 0.7606                  & 0.7514                                                                      & 0.7885                  & 0.7627                   \\
PLE+DML                & $\boldsymbol{0.8151}^*$ & $\boldsymbol{0.7533}^*$ & ${0.7631}^*$                                                              & ${0.7893}^*$          & ${0.7640}^*$           \\
\hline
\end{tabular}
\end{table}
%\vspace{-0.2cm}
\subsection{Experimental Settings for Public Data}
%\vspace{-0.1cm}
\subsubsection{Datasets} 
We evaluate our methods on two public datasets.
%\vspace{-0.07cm}
\begin{itemize}[leftmargin=.1in]
\item \textbf{MovieLens-1M}\cite{harper2015movielens}: One of the currently released MovieLens datasets, which contains 1 million movie ratings from 6,040 users on 3,416 movies.
\item \textbf{Amazon}\cite{mcauley2015image}: A series of datasets consisting of product reviews from Amazon.com. We use the sub-category of "Electronics"  including 1.7 million reviews from 192,403 users on 63,001 items.
\end{itemize}
For ML-1M, we introduce the binary classification task of positive rating prediction (>=4) and the regression task of rating estimation. These two tasks are strictly correlated. For electronics, following \cite{wang2022can}, we first augment the dataset by randomly sampling un-rated items for every user. Moreover, we make sure the number of the un-rated items is the same as the number of the rated items for each user. Furthermore, we introduce two binary classification tasks, namely rating prediction (whether a rating exists) and positive rating prediction. Compared with the tasks of ML-1M, the negative transfer is more likely to occur as the task relationship here is more complex(The pearson correlation coefficient \cite{wiki:Pearson_correlation_coefficient} of these two labels is around 0.7). Both data are randomly split into the training set, validation set, and test set by the ratio of 8:1:1.  

%\vspace{-0.2cm}
\subsubsection{Evaluation Metrics}
 The merge function $\Phi$ in Equation \eqref{eq:score_merge} assumes that the model can estimate accurate interaction probabilities for binary classification tasks (e.g. clicking) and absolute values for regression tasks (e.g. watch time). Therefore, instead of the ranking metrics, such as NDCG \cite{wang2013theoretical} and MRR \cite{enwiki:1095286224}, we use the metrics of AUC \cite{flach2011coherent} for classification tasks and Mean Squared Error (MSE) \cite{enwiki:1127519968} for regression tasks. Please note that many other recommendation literature, such as \cite{tang2020progressive,zhao2019recommending,ding2021mssm}, also use similar metrics. For AUC, a bigger value indicates better performance. While, for MSE, it is the smaller the better.
%\vspace{-0.3cm}
\subsubsection{Models}
As \textbf{\textit{soft parameter sharing}} methods need resources to store and train multiple sets of parameters, they are not widely applied in recommender systems. Thus, we select base models to cover the other three categories. The models include Shared-Bottom(SB)\cite{caruana1997multitask}, MSSM\cite{ding2021mssm}, MMOE\cite{ma2018modeling}, and PLE\cite{tang2020progressive}. MSSM is a recent method belonging to the \textbf{\textit{Customized Routing}} category and achieves better results than SNR\cite{ma2019snr} and Cross-Stitch\cite{misra2016cross}. Though with the same category of \textbf{\textit{Dynamic Gating}}, both MMOE and PLE are tested owing to their popularity. For each base model, we will verify whether DML can achieve additional gains. For reference, we also provide the performance of single task models.
%\vspace{-0.3cm}
\subsubsection{Implementation Details}
For each feature, we use the embedding size 8. As suggested by the original papers, we use 1 level bottom sub-networks for MMOE, MSSM, and SB while 2 levels for PLE. For SB, a sub-network of 1 layer structure with 128 output dimensions is shared by the tasks. For other multi-task models, each bottom level includes three sub-networks, which have the same aforementioned architecture. For MSSM and PLE, task-specific and shared sub-networks are designated. For multi-task models, each task tower is of the three layers MLP structure (128,80,1) and each task is assigned equal loss weight. For the single-task model, each task utilizes the four layers MLP structure (128,128,80,1). For the first two MLPs at Figure \ref{fig:dml:gkd}, we utilize the one layer structure with 80 as the output dimension. For the last MLP at Figure \ref{fig:dml:gkd}, a one layer structure with 1 as the output size is used. If not explicitly specified, RELU \cite{DBLP:conf/iclr/AroraBMM18} is used as the default activation function. All models are implemented with tensorflow \cite{abadi2016tensorflow} and optimized using the Adam \cite{DBLP:journals/corr/KingmaB14} optimizer  with learning rate 0.001 and mini-batch size 512. We run 20 times for each test to report the results.
%\vspace{-0.3cm}

\begin{table}
%\footnotesize
\normalsize
\centering
\caption{Further Analysis Results}
\label{tab:analysis}
\begin{tabular}{lcc|cc} 
\hline
\multirow{2}{*}{Model}                          & \multicolumn{2}{c|}{ML-1M}                              & \multicolumn{2}{c}{Electronics}                          \\ 
\cline{2-5}
                                                & $AUC$                      & $MSE$                      & $AUC_{rate}$                    & $AUC_{pos}$                     \\ 
\hline\hline
$MSSM$                                          & 0.8128                     & 0.7651                     & 0.7883                     & 0.7627                      \\
$MSSM+CTFM$                                     & 0.8134                     & 0.7654                     & 0.7891                     & 0.7636                      \\
$MSSM+GKD$                                      & 0.8129                     & 0.7636                     & 0.7886                     & 0.7633                      \\
$MSSM+DML_{v0}$                                 & 0.8138 & 0.7625 & 0.7884 & 0.7631  \\
$MSSM+DML$                                      & \textbf{0.8141}            & \textbf{0.7611}            & \textbf{0.7892}            & \textbf{0.7641}             \\ 
\hline
\begin{tabular}[c]{@{}l@{}}$PLE$\\\end{tabular} & 0.8122                     & 0.7606                     & 0.7885                     & 0.7627                      \\
$PLE+CTFM$                                      & 0.8133                     & 0.7603                     & 0.7890                     & 0.7633                      \\
$PLE+GKD$                                       & 0.8137 & 0.7574 & 0.7887 & 0.7634  \\
$PLE+DML_{v0}$                                  & 0.8141                     & 0.7536                     & 0.7886                     & 0.7629                      \\
$PLE+DML$                                       & \textbf{0.8151}            & \textbf{0.7533}            & \textbf{0.7893}            & \textbf{0.7640}             \\
\hline
\end{tabular}
\end{table}
\subsection{Overall Performance for Public Data}
Please refer to Table \ref{tab:overview} for the overall results. First, DML achieves significant gains across all these tested multi-task models on the two public datasets, which shows DML's wide compatibility. Second, DML-enhanced PLE and MMOE get the best performance for MovieLens and Electronics respectively. Considering their wide application in recommender systems, the results are as expected. Third, the multi-task models perform better than the single task models thanks to the knowledge transfer between tasks.   

Besides AUC and MSE, DML should help to foster task consistency with \textit{CTFM} and \textit{GKD}. As the tasks of MovieLens are rigorously correlated, we verify whether DML really enhances task consistency on this data. First, we construct pairs of samples with different rating scores and count the pair numbers. Second, we count the number of pairs, for which the prediction scores of both tasks are in the same pair order as the rating score. The enhancement of the pair order consistency among the two prediction scores and rating score should positively contribute to the performance. Then, we can compute the metric of 'Consistency Ratio'. The listed data in Table \ref{tab:overview} agree with our anticipation. (For Shared-Bottom, we also observe more pairs, for which predictions of both tasks are in rating score's reverse order. This can explain its worse performance in spite of the better consistency ratio.)   
%\vspace{-0.3cm}
\subsection{Further Analysis on Public Data}
We select the two latest algorithms of PLE\cite{tang2020progressive} and MSSM\cite{ding2021mssm} to appraise the value of $DML$'s components, namely $CTFM$ and $GKD$. Without the stop gradient operation, $CTFM$ will be very similar to the common attention mechanism. To prove the benefit of $CTFM$'s design, we also add the assessment for $DML_{v0}$, which reserve the design of $GKD$ while remove the gradient blocking operation of $CTFM$. Please refer to Table \ref{tab:analysis} for the evaluation results. First, $CTFM$ and $GKD$ both contribute considerable gains over the base model. Second, as the integrated model, $DML$ enhances the performance further. Third, $DML_{v0}$ is consistently worse than $DML$, which corroborates the value of reserving task-awareness. Compared with $CTFM$ and $GKD$, $DML_{v0}$ performs better on MovieLens while much worse on Electronics. The task relationship of Electronics is more complex and negative transfer across tasks usually exhibits more severe impact due to task conflicts. In this case, compared with vanilla attention, $CTFM$ obtains substantial gains. 
%\vspace{-0.3cm}
\subsection{Online A/B Testing}
DML is applied to the ranking stage\cite{covington2016deep} of an industrial large-scale news recommender system. PLE \cite{tang2020progressive} is utilized as the base model. And the main prediction tasks are the binary classification task of Click Through Rate (CTR) and the regression task of item watch time. First, after the model converge by training with billions of samples, the AUC metric for CTR consistently increases 0.12\% and the MSE metric for watch time decreases 0.14\%. Moreover, the most important online metrics include effective PV(count of Page Views with watch time exceeding a threshold) and total watch time. We randomly distributed online users to two buckets with the base PLE model or PLE+DML model and evaluated the performance for two weeks. DML achieves significant (p<0.05) gains over the base model by 1.22\% for effective PV and 0.61\% for total watch time. DML has been deployed to our online environment based on the results.

%\vspace{-0.3cm}
\section{Conclusion}
In this papaer, we propose the framework of \textbf{D}eep \textbf{M}utual \textbf{L}earning across task towers(\textbf{DML}), which is compatible with various backbone multi-task networks. Extensive offline experiments help to verify DML's effectiveness on multiple real-world datasets and across various base models. Moreover, thorough ablation studies are carried out to verify and understand the value of each newly introduced module. Finally, DML achieves significant online gains and has already been deployed to the online platform.

%%\section{CITE}

%\clearpage
\bibliographystyle{ACM-Reference-Format}
\balance
\bibliography{reference}

\end{document}